\documentclass[sn-mathphys,iicol,Numbered]{sn-jnl}

\usepackage{graphicx}%
\usepackage{multirow}%
\usepackage{amsmath,amssymb,amsfonts}%
\usepackage{amsthm}%
\usepackage{mathrsfs}%
\usepackage[title]{appendix}%
\usepackage{xcolor}%
\usepackage{textcomp}%
\usepackage{manyfoot}%
\usepackage{booktabs}%
\usepackage{algorithm}%
\usepackage{algorithmicx}%
\usepackage{algpseudocode}%
\usepackage{listings}%
\usepackage{makecell} 
\usepackage{xspace}
\usepackage{comment}
\usepackage{sidecap}
\usepackage{siunitx}
\usepackage{orcidlink}
\usepackage{lineno}

\newcommand{\GeantFour}{{\scshape Geant4}\xspace}
\newcommand{\DPMJet}{{\scshape DPMJet}\xspace}
\newcommand{\Fluka}{{\scshape Fluka}\xspace}

\newcommand{\SND} {\mbox{SND@LHC}\xspace}

\begin{document}

\title[Article Title]{Measurement of the muon flux at the SND@LHC experiment}


\author[1,2]{\fnm{R.}\sur{Albanese}~\orcidlink{0000-0003-4586-8068}}
\author[1]{\fnm{A.}\sur{Alexandrov}~\orcidlink{0000-0002-1813-1485}}
\author[1,2]{\fnm{F.}\sur{Alicante}~\orcidlink{0009-0003-3240-830X}}
\author[3]{\fnm{A.}\sur{Anokhina}~\orcidlink{0000-0002-4654-4535}}
\author[1,2]{\fnm{T.}\sur{Asada}~\orcidlink{0000-0002-2482-8289}}
\author[4,5]{\fnm{C.}\sur{Battilana}~\orcidlink{0000-0002-3753-3068}}
\author[6]{\fnm{A.}\sur{Bay}~\orcidlink{0000-0002-4862-9399}}
\author[7]{\fnm{C.}\sur{Betancourt}~\orcidlink{0000-0001-9886-7427}}
\author[8]{\fnm{D.}\sur{Bick}~\orcidlink{0000-0001-5657-8248}}
\author[9]{\fnm{R.}\sur{Biswas}~\orcidlink{0009-0005-7034-6706}}
% ...
\affil[1]{\orgname{Sezione INFN di Napoli},\orgaddress{\city{Napoli},\postcode{80126},\country{Italy}}}
\affil[2]{\orgname{Universit\`{a} di Napoli ``Federico II''},\orgaddress{\city{Napoli},\postcode{80126},\country{Italy}}}
\affil[3]{\orgname{Affiliated with an institute covered by a cooperation agreement with CERN}}
\affil[4]{\orgname{Sezione INFN di Bologna},\orgaddress{\city{Bologna},\postcode{40127},\country{Italy}}}
\affil[5]{\orgname{Universit\`{a} di Bologna},\orgaddress{\city{Bologna},\postcode{40127},\country{Italy}}}
\affil[6]{\orgname{Institute of Physics, EPFL},\orgaddress{\city{Lausanne},\postcode{1015},\country{Switzerland}}}
\affil[7]{\orgname{Physik-Institut, UZH},\orgaddress{\city{Z\"{u}rich},\postcode{8057},\country{Switzerland}}}
\affil[8]{\orgname{Hamburg University},\orgaddress{\city{Hamburg},\postcode{22761},\country{Germany}}}
\affil[9]{\orgname{European Organization for Nuclear Research (CERN)},\orgaddress{\city{Geneva},\postcode{1211},\country{Switzerland}}}
% ...







\abstract{The Scattering and Neutrino Detector at the LHC (\SND)  started taking data at the beginning of Run 3 of the LHC. The experiment is designed to perform measurements with neutrinos produced in proton-proton collisions at the LHC in an energy range between~100~GeV and~1~TeV. It covers a previously unexplored pseudo-rapidity range of \mbox{$7.2<\eta<8.4$}. The detector is located 480~m downstream of the ATLAS interaction point in the TI18 tunnel. It comprises a veto system, a target consisting of tungsten plates interleaved with nuclear emulsion and scintillating fiber (SciFi) trackers, followed by a muon detector (UpStream, US and DownStream, DS). In this article we report the measurement of the muon flux in three subdetectors:  the emulsion, the SciFi trackers and the DownStream Muon detector. 

The muon flux per integrated luminosity through an \mbox{18$\times$18 cm$^{2}$} area in the emulsion is:\\
\centerline{\mbox{$1.5 \pm 0.1(\textrm{stat}) \times 10^4\,\textrm{fb/cm}^{2}$}}. The muon flux per integrated luminosity through a \mbox{31$\times$31 cm$^{2}$} area in the centre of the SciFi is:\\
\centerline{\mbox{$2.06\pm0.01(\textrm{stat})\pm0.12(\textrm{sys}) \times 10^{4} \textrm{fb/cm}^{2}$}} 
The muon flux per integrated luminosity through a \mbox{52$\times$52 cm$^{2}$} area in the centre of the downstream muon system is:\\
\centerline{\mbox{$2.35\pm0.01(\textrm{stat})\pm0.10(\textrm{sys}) \times 10^{4}\,\textrm{fb/cm}^{2}$}}  The total relative uncertainty of the measurements by the electronic detectors  is 6~$\%$ for the SciFi and 4~$\%$ for the DS measurement. The Monte Carlo simulation prediction of these fluxes is~20-25~$\%$ lower than the measured values.}

\keywords{muon flux, SND@LHC, LHC, emulsion, scintillating fibres}



\maketitle
\section{Introduction}\label{sec:introduction}

The \SND detector~\cite{Acampora:2834502} is designed to perform measurements with high energy neutrinos \mbox{(100~GeV--1~TeV)} produced in proton-proton collisions at the LHC in the forward pseudo-rapidity region \mbox{$7.2 < \eta < 8.4$}. It is a compact, stand-alone experiment located 480~m away from the ATLAS interaction point (IP1) in the TI18 tunnel, where it is shielded from collision debris by around 100~m of rock and concrete. The signal events for the experiment are neutrino interactions~\cite{Albanese:031802} and searches for dark matter scatterings. However, the majority of recorded events consists of muons arriving from the particles produced in proton-proton collisions at IP1. Since these muons are the main source of background for the neutrino search, it was necessary to do a measurement of the muon flux in the \SND detector.

\section{Detector}\label{sec:detector}

\begin{figure*}[ht]%
\centering
\includegraphics[width=1.0\textwidth]{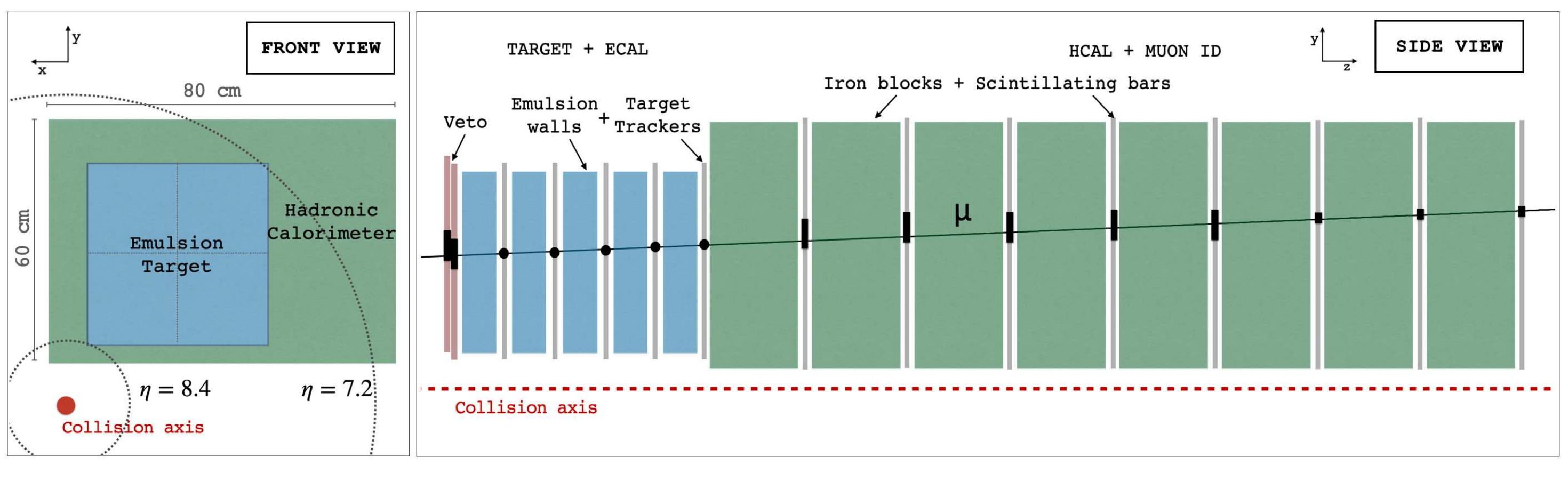}
\caption{Schematic layout of the \SND detector. The pseudo-rapidity $\eta$ values are the limits for particles hitting the lower left and the upper right corner of the ECC. The side view includes an illustration of a passing-through muon with hits in all subdetectors. For a view of the $x-y$ ranges covered by the SciFi and the muon DS system see Figures~\ref{fig:Sf11_data_muon_xy}~and~\ref{fig:DS13_data_muon_xy}.}\label{fig:detector_layout}
\end{figure*}

Figure~\ref{fig:detector_layout} shows the \SND detector. The electronic detectors provide the time stamp of the neutrino interaction, preselect the interaction region while the vertex is reconstructed using tracks in the emulsion. The veto system  is used to tag muons and other charged particles entering the detector from the IP1 direction.

The veto system comprises two parallel planes of scintillating bars. Each plane consists of seven 1 $\times$ 6 $\times$ 42~cm$^{3}$ vertically stacked bars of plastic scintillator.

The target section contains five walls. Each wall consists of four units ('bricks') of Emulsion Cloud Chambers (ECC) and is followed by a scintillating fiber (SciFi) station for tracking. 

Each SciFi station consists of one horizontal and one vertical 39~$\times$~39~cm$^{2}$ plane. Each plane comprises six  staggered layers of 250~\textmu m diameter polystyrene-based scintillating fibers. The single particle spatial resolution in one plane is  $\sim$150~\textmu m and the time resolution for a particle crossing both $x$ and $y$ planes is about 250~ps.

The muon system consists of two parts:  the first five stations (UpStream, US), and the last three stations (DownsStream (DS), see Figure~\ref{fig:detector_layout}). Each US station consists of 10 stacked horizontal scintillator bars of 82.5~$\times$~6~$\times$~1~cm$^{3}$, resulting in a coarse $y$ view. A DS station consists of two layers of thinner bars measuring 82.5~$\times$~1~$\times$~1~cm$^{3}$, arranged in alternating $x$ and $y$ planes, allowing for a spatial resolution in each axis of less than 1~cm. The eight scintillator stations are interleaved with 20~cm thick iron blocks.  Events with hits in the DS detector and the SciFi tracker are used to identify muons.

A right-handed coordinate system is used, with $z$ along the nominal proton-proton collision axis and pointing away from IP1, $x$ pointing away from the centre of the LHC, and $y$ vertically aligned and pointing upwards.

All signals exceeding preset thresholds are read out by the front-end electronics and clustered in time to form events. A software noise filter is applied to the events online, resulting in negligible detector deadtime and negligible loss in signal efficiency. Events satisfying certain topological criteria, such as the presence of hits in several detector planes, are read out. At the highest instantaneous luminosity in 2022 (2.5~$\times$~10$^{34}$~cm$^{-2}$\,s$^{-1}$) this generated a rate of around 5.4~kHz.

\section{Data and Monte Carlo simulations}\label{samples}

\subsection{Data sample}\label{subsec:data}
\subsubsection{Emulsion}
The data used in the analysis of the emulsion was from a brick that was irradiated during the (LHC commissioning) period  \mbox{7$^{\textrm{th}}$ May} - \mbox{26$^{\textrm{th}}$ July 2022}.  The integrated luminosity for this period was 0.5~fb$^{-1}$.
\subsubsection{Electronic detectors}
During the production 13.6~TeV~proton  physics period in 2022 \SND recorded an integrated luminosity of 36.8~fb$^{-1}$. This amounts to $95\%$ of the total 38.7~fb$^{-1}$ delivered luminosity at IP1, as reported by ATLAS~\cite{ATL-DAPR-PUB-2023-001}. We used two runs (see \mbox{Table~\ref{tab:selected_runs}}) from this data for the muon flux measurement using the electronic detectors.

\begin{table*}
    \caption{List of the two selected SND@LHC~2022 data runs. The runs are chosen to have large event counts, high delivered luminosity, isolated LHC bunches of Beam~2 passing without collisions, and different LHC filling schemes.  }\label{tab:selected_runs}
    \begin{tabular}{c|c|c|c|c|c|c}
    \hline\hline
        \makecell{LHC fill\\ number} & \makecell{ integrated\\luminosity\\$[$fb$^{-1}]$ } & \makecell{mean number of\\interactions\\ at IP1 per\\ bunch crossing} & \makecell{SND@LHC \\ run number}& \makecell{recorded events\\ by~\SND\\ $[10^{6}]$} & \makecell{date, \\year 2022} & \makecell{duration \\ $[$h$]$} \\
        \hline
         8088 & 0.337 & 35.2 & 4705 & 71 & 3 Aug  & 12.5 \\
         8297 & 0.529 & 45.4 &  5086 & 101 & 20 Oct  & 19.8 \\
         \hline\hline
    \end{tabular}
\end{table*}
\subsubsection{Muons not from beams colliding in IP1}
For the calculation of the muon flux, we are only interested in the muons that come from $pp$ collisions in IP1.
The LHC filling scheme specifies which bunches cross at different interaction points and which bunches of Beam~1 and Beam~2 are circulating in the LHC without colliding.\footnote{The clockwise circulating beam is denoted Beam~1, while the counter clockwise circulating beam is denoted Beam~2~\cite{Brüning:782076}.}  Since the  \SND detector is 480~m away from IP1, there is a phase shift between the  filling scheme and the \SND event timestamp. The phase adjustments for both beams are determined by finding the maximum overlap with \SND event rates. The synchronized bunch structure then allows us to determine whether an event is associated with a collision at IP1, whether it originates from  Beam~1 without colliding in IP1,  or whether it  originates from Beam~2 without colliding in IP2. For the determination of the muon flux, the latter two contributions have to be subtracted from the total number of recorded muons associated with IP1 collisions~\cite{Ilieva:2859193}. The muon contribution from IP2 collisions is negligible given the small difference in the event rates associated with circulation of non-colliding Beam~2 bunches and IP2 collisions.

\subsection{Monte Carlo Simulations}\label{subsec:mc}
The $pp$ event generation was done with \DPMJet (Dual Parton Model)~\cite{Fedynitch:2015kcn}. 
The subsequent muon production from the $pp$ collisions  was simulated with \Fluka\cite{ Ahdida:2022gjl}.  The propagation of collision debris in the LHC towards the \SND detector was done using the LHC \Fluka model~\cite{Prelipcean:2839993}. The particle transport was stopped at a 1.8$\times$1.8~m$^{2}$ scoring plane, located in the rock about 60~m upstream of \SND. This dataset consists of particles from 200~$\times$~10$^{6}$~$pp$ collisions simulated with LHC Run~3 beam conditions. 

The propagation of muons from the scoring plane to the detector and their interactions were modelled with a \GeantFour simulation~\cite{GEANT4:2002zbu} of \SND and its surroundings.

\section{Track finding and fitting methods for the electronic detectors}\label{sec:tracking}

Most muons traveling from IP1 towards \SND leave straight tracks in the detector. Two track-finding methods were implemented in the \SND software framework. One of them makes use of a custom track-finding solution that minimizes the residuals between measured points and a straight-line track candidate, denoted Simple Tracking~(ST). The other tracking approach employs the Hough transform~\mbox{\cite{osti_4746348}} pattern recognition method and is referred to as Hough Transform~(HT). In both cases, the track fitting is done using the Kalman filter method in the GENFIT package~\cite{ HOPPNER2010518}. 

Since the SciFi and DS detectors have a different granularity and the acceptance of the DS is 2.4~times larger than that of the SciFi, 
the muon flux is determined independently in these two detector subsystems. 

Track building in these subsystems is done separately in the horizontal~\mbox{$x-z$} and vertical~\mbox{$y-z$} plane. The final three dimensional~track is built by combining the horizontal and vertical tracks~\cite{Ilieva:2859193}.

The tracking efficiency in each detector and for each of the tracking methods is estimated using data (see Table~\ref{tab:trackingEff_data}). The uncertainty of the efficiency is evaluated as three times the standard deviation of tracking efficiency values over 1$\times$1~cm$^{2}~x-y$ detector coordinate bins. The HT method has a better efficiency and it is used as the baseline for this analysis.
\begin{table}[t]
    \caption{SciFi and DS tracking efficiencies for simple and Hough transform tracking methods.}
    \label{tab:trackingEff_data}
    \centering
    \begin{tabular}{c|c|c}
        \hline \hline
         system & tracking algorithm & tracking efficiency\\
         \hline
         SciFi & \makecell{simple tracking\\Hough transform} & \makecell{$0.868\pm0.009 $\\ $0.956\pm0.007 $ }\\
         DS & \makecell{simple tracking\\Hough transform} & \makecell{$0.937\pm0.007 $\\ $0.944\pm0.009 $ }\\
         \hline \hline
    \end{tabular}
\end{table}

\section{Angular distribution in the electronic detectors}\label{sec:slopes}

\begin{figure}[t]
\centering
    \includegraphics[width=0.37\textwidth]{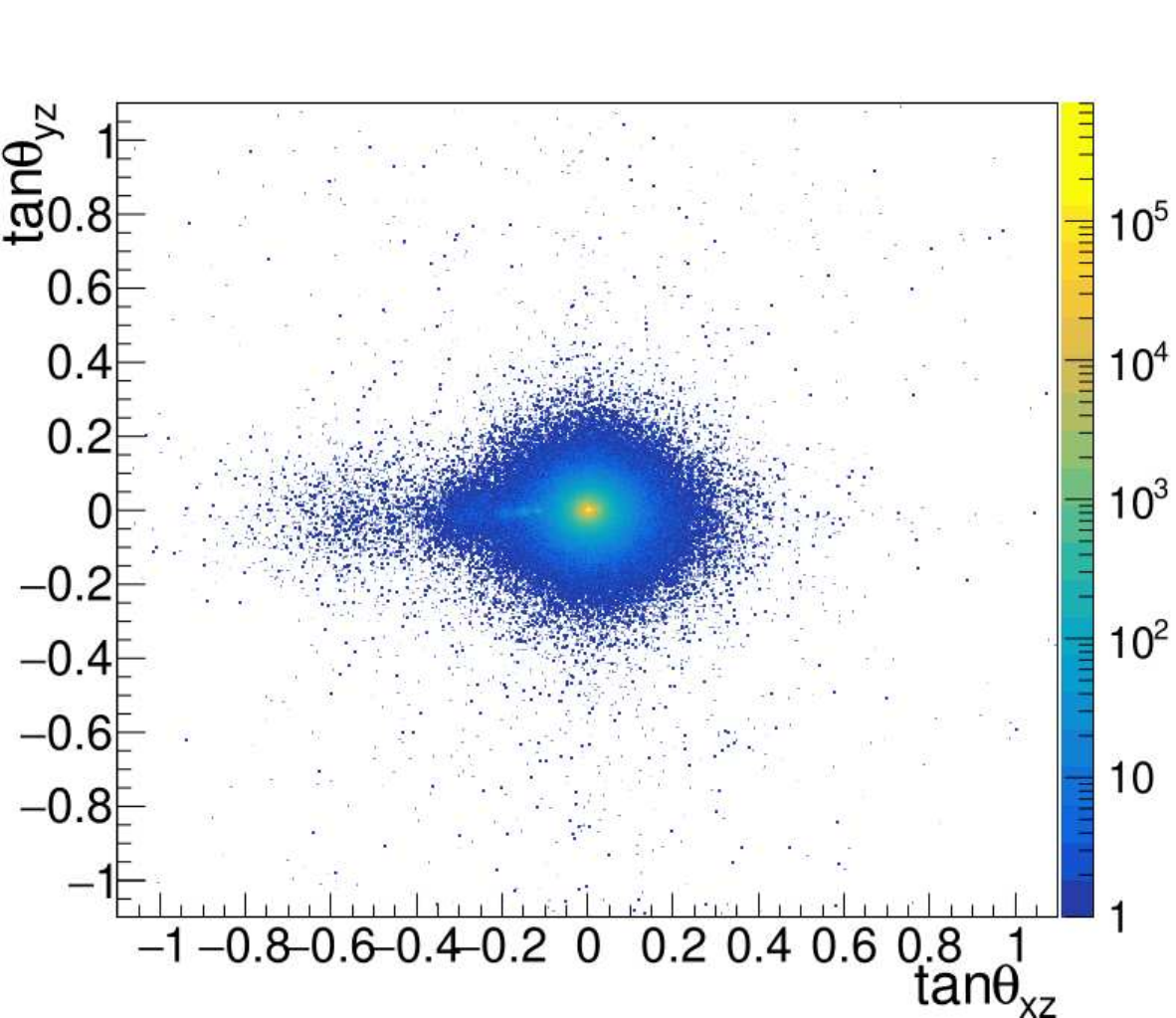}
\caption{SciFi track slopes of reconstructed tracks. The slopes in the horizontal $x-z$ plane ($\tan \theta_{\mathrm{xz}}$) and in the vertical $y-z$ plane ($\tan \theta_{\mathrm{yz}}$) are derived from the differences of the track coordinates between the first and the last track point in the detector.}\label{fig:Sf11_2Dslopes}
\end{figure}

\begin{figure}[t]
    \centering
    \includegraphics[width=0.37\textwidth]{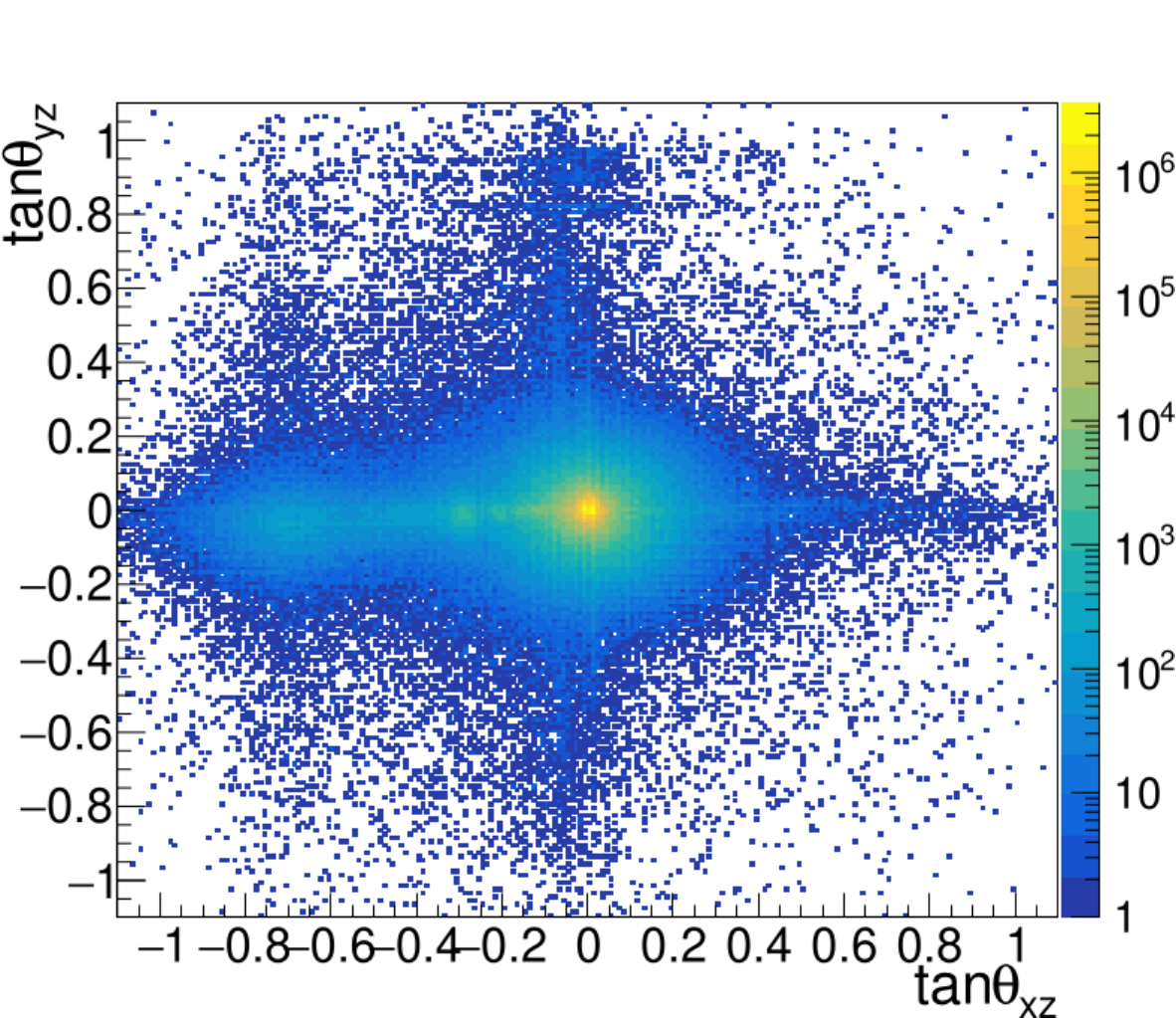}\\
    \caption{DS track slopes of reconstructed tracks. The slopes in the horizontal $x-z$ plane ($\tan \theta_{\mathrm{xz}}$) and in the vertical $y-z$ plane ($\tan \theta_{\mathrm{yz}}$) are derived from the differences of the track coordinates between the first and the last track point in the detector. }
    \label{fig:DS13_2Dslopes}
\end{figure}

\begin{figure}[htbp]
\centering
\includegraphics[width=0.37\textwidth]{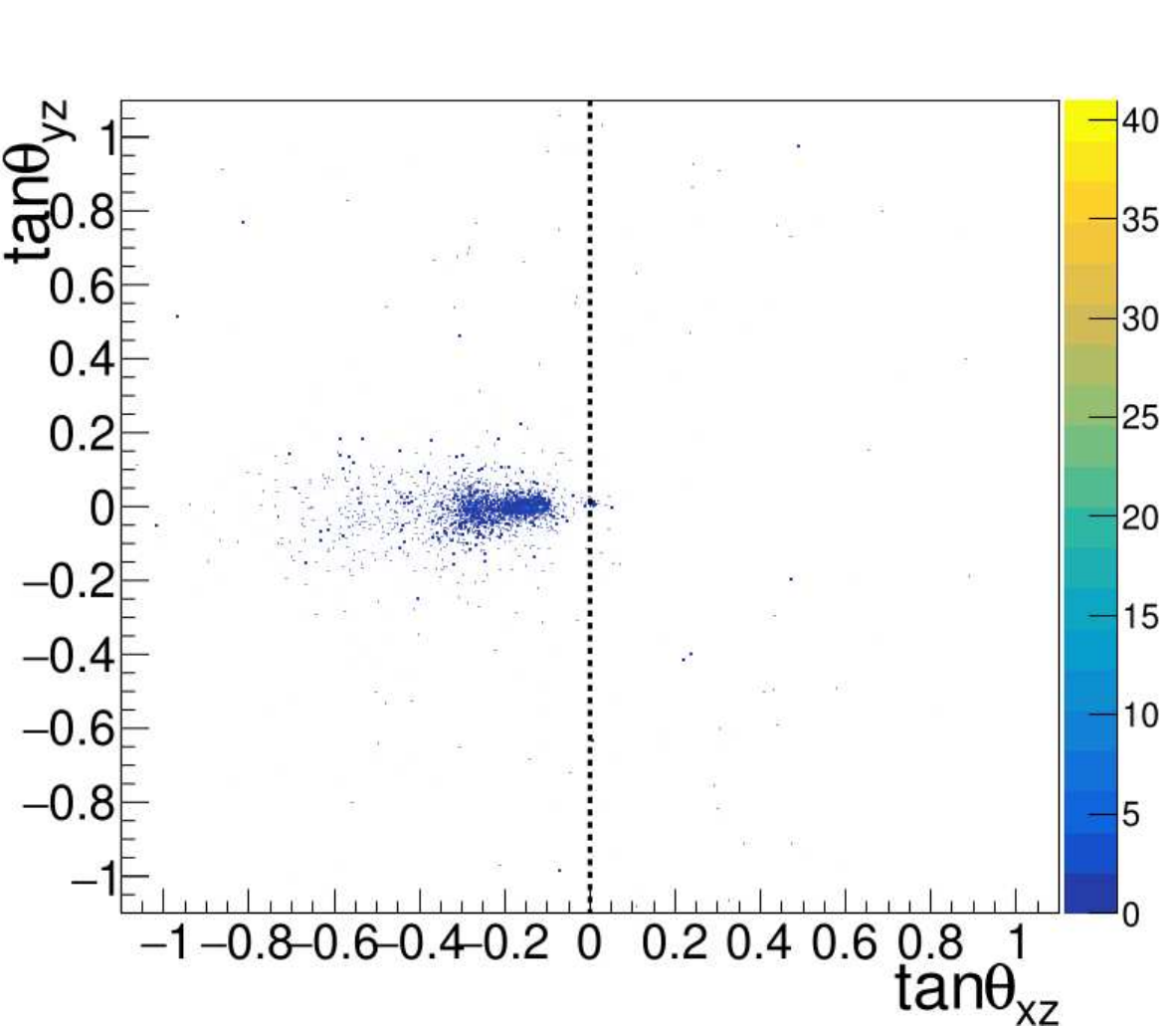}
\caption{SciFi track slopes for data events in sync with B2noB1 LHC bunches.}\label{fig:2Dslopes_b2nob1}
\end{figure}

\begin{figure}[htbp]
    \centering
    \includegraphics[width=0.37\textwidth]{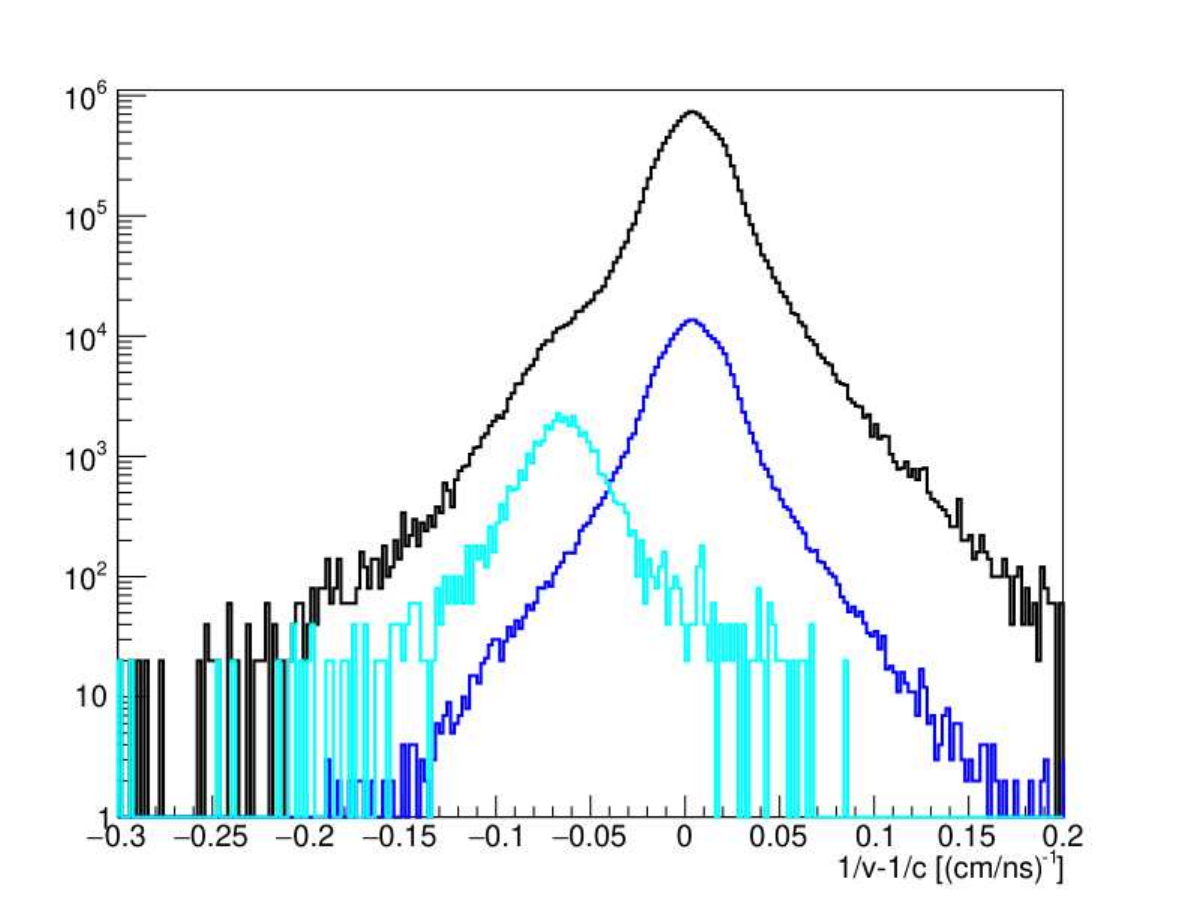}
    \caption{Track separation by particle propagation direction using SciFi tracks. The black curve corresponds to tracks reconstructed in all events. The blue curve corresponds to tracks in events of non-colliding Beam~1 bunches and no present bunches of Beam~2~(B1noB2). The cyan curve is for events of non-colliding Beam~2 bunches and no present bunches of Beam~1~(B2noB1) bunches. Tracks of particles moving from the DS towards the SciFi (backward going tracks), must have \mbox{$1/v\,-\,1/c$} values around $-2/c$ or -0.067~(cm/ns)$^{-1}$. For each track, the value $1/v$ is obtained from a straight line fit to a plot of position difference versus timing difference between each track point and its first upstream point in the SciFi detector.}
    \label{fig:track_direction}
\end{figure}

Most muons reaching \SND have tracks with small angles with respect to the $z$ axis, see Figures~\ref{fig:Sf11_2Dslopes}~and~\ref{fig:DS13_2Dslopes}. The main peak corresponds to the source at IP1. This peak has large tails due to multiple scattering along the 480~m path from IP1 to \SND. The structures at negative slopes originate from beam-gas interactions. As shown in Figure~\ref{fig:2Dslopes_b2nob1}, after selecting \SND events  corresponding to non-colliding Beam~2 bunches and no present bunches of Beam~1~(B2noB1), almost all reconstructed tracks have negative $x-z$ slopes. Track direction studies based on detector hit timing show that particles with reconstructed tracks in such B2noB1~events enter the detector from the back (see Figure~\ref{fig:track_direction}). Therefore, the origin of these particles must be downstream of the DS stations. 


\begin{figure}[htbp]
    \centering
     \includegraphics[width=0.37\textwidth]{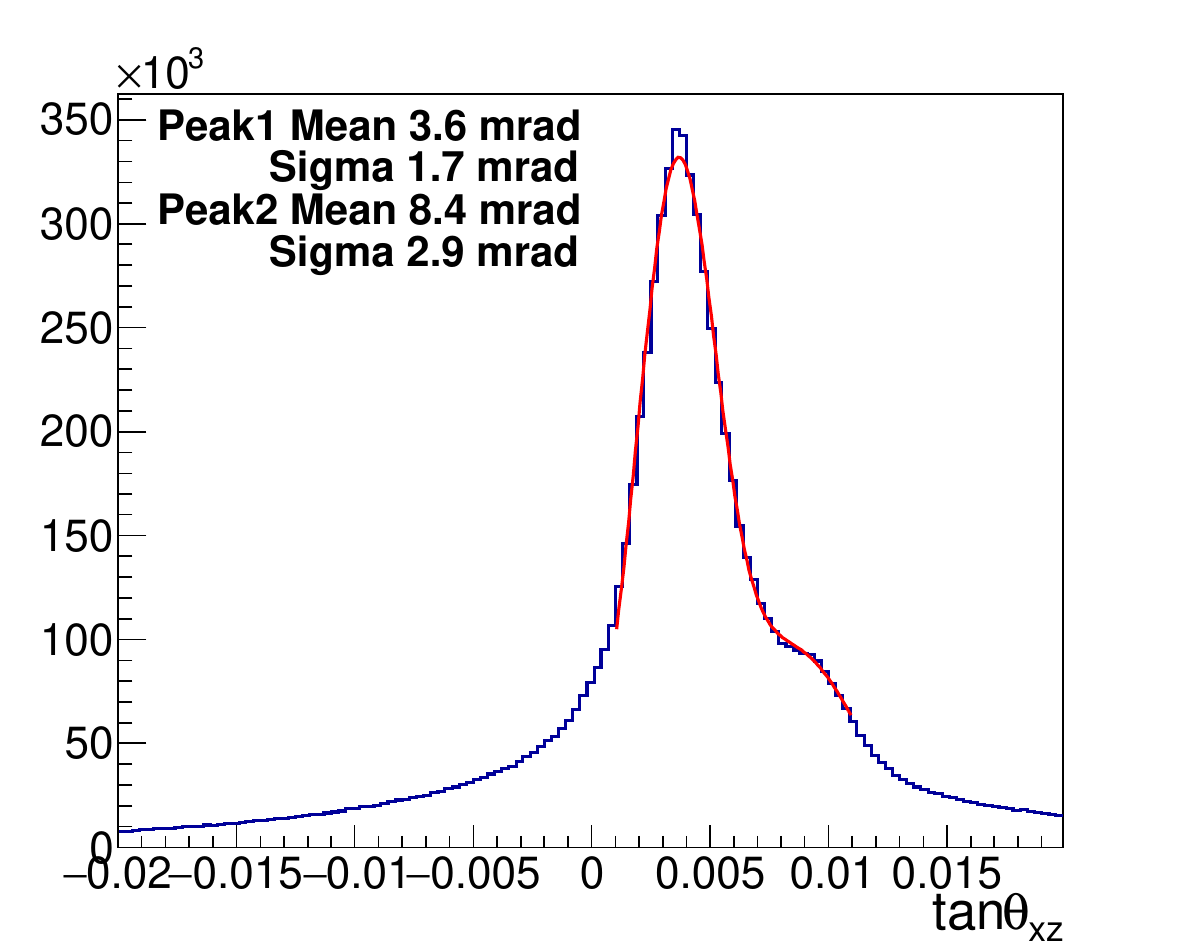}
    \caption{Angles of SciFi tracks in the $x-z$ plane. The two peaks are fitted with a two Gaussian function. }
    \label{fig:central_peaks}
\end{figure}

Figure~\ref{fig:central_peaks} shows the angles of SciFi tracks in the $x-z$ plane in the range of very small angles ($-0.02< \tan{\theta_{\mathrm{xz}}}<0.02$).  The angular distance between the two slightly shifted peaks is about 5~mrad. A similar structure is seen in the emulsion data (see Figure~\ref{fig:track_slopes}) and the Monte Carlo simulation (see Figure~\ref{fig:MC_central_peaks}). From the \Fluka  simulation  it is known that muons in the main peak originate  at IP1 and muons in the other peak are from particle (pion and kaon) decays at various locations~\cite{Ilieva:2859193}. 

\section{Muon flux in the electronic detectors}\label{sec:flux}

The muon flux is defined as the number of reconstructed tracks per corresponding IP1 integrated luminosity and unit detector area. The number of tracks is corrected for the tracking efficiency.  

\begin{figure}[htbp]
    \centering
    \includegraphics[width=0.37\textwidth]{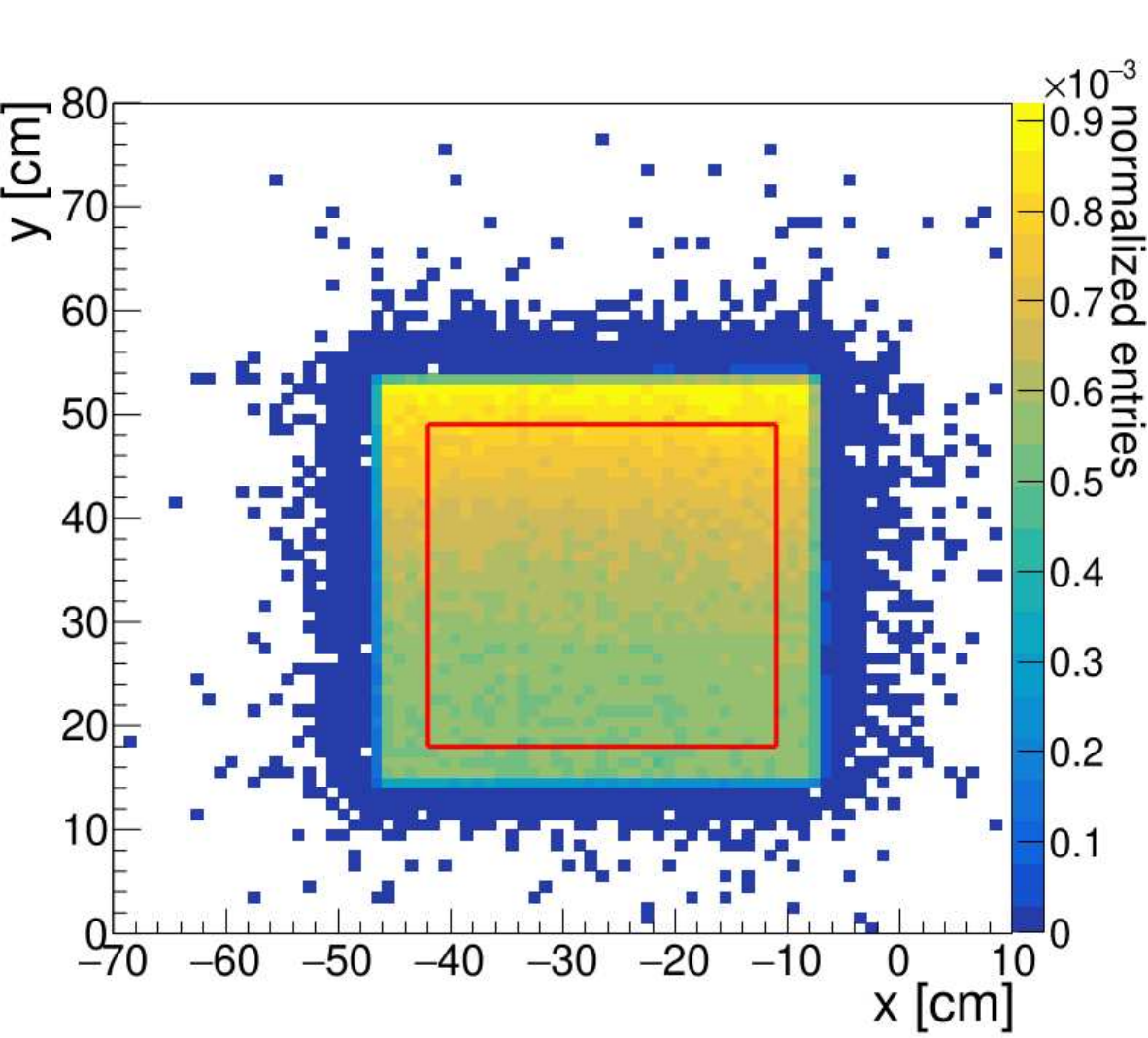}\\
    \caption{Distribution of SciFi tracks at the most upstream detector plane. The distribution is normalized to unit integral. The red border delimits the region considered for the SciFi muon flux measurement.} \label{fig:Sf11_data_muon_xy}
\end{figure}

\begin{figure}[htbp]
    \centering
    \includegraphics[width=0.37\textwidth]{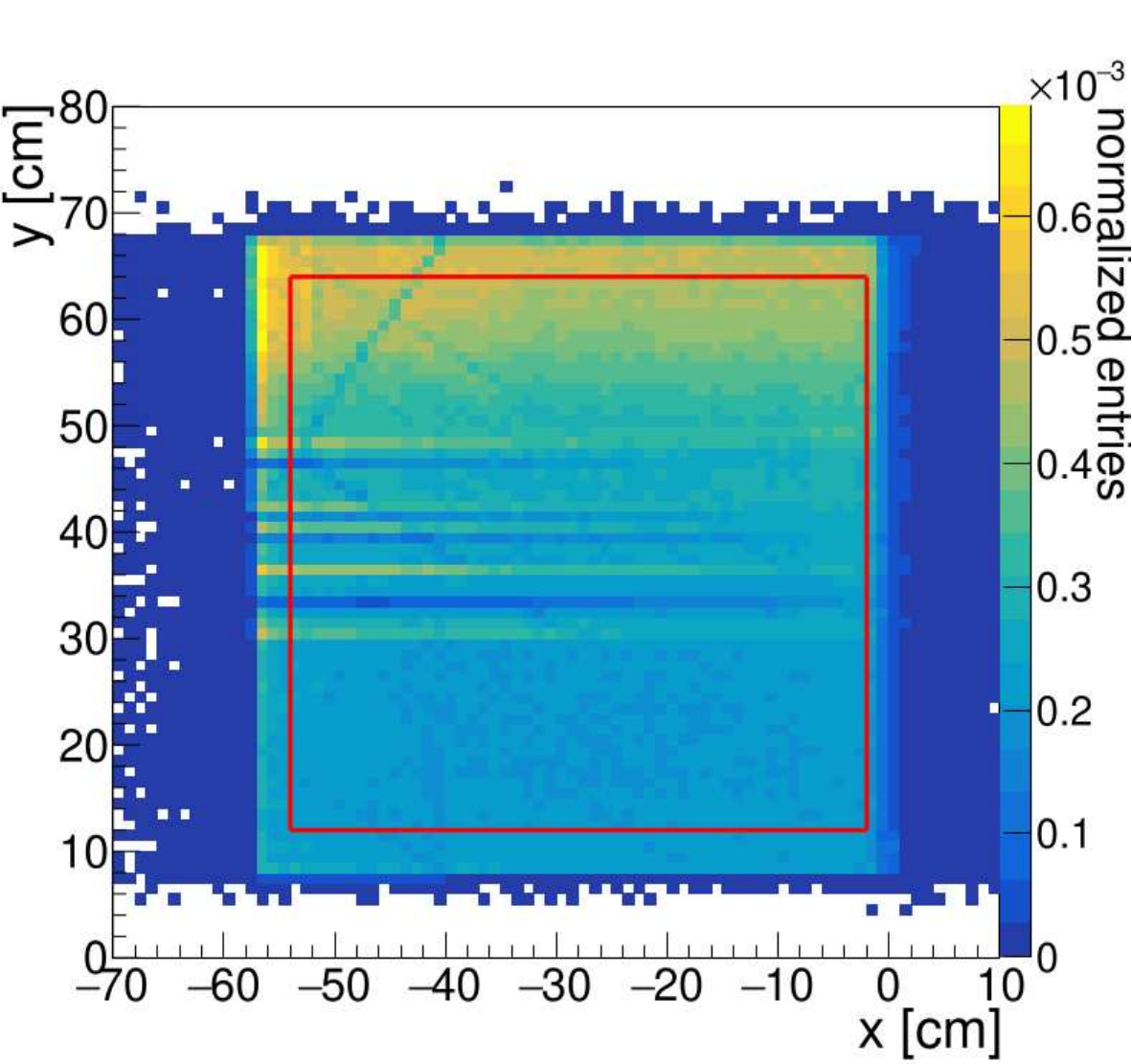}\\
    \caption{Distribution of DS tracks at the most upstream detector plane. The distribution is normalized to unit integral. Horizontal stripes of lower counts in the central part of the detector are caused by scintillator bar inefficiencies. The red border delimits the region considered for the DS muon flux measurement.}\label{fig:DS13_data_muon_xy}
\end{figure}

The muon flux in the SciFi and DS detectors is estimated in an area with uniform tracking efficiency~\cite{Ilieva:2859193}. For the SciFi this is the area between \mbox{$-42\,\textrm{cm}\leq x \leq-11\,\textrm{cm}$} and \mbox{$18\,\textrm{cm}\leq y \leq49\,\textrm{cm}$} (31$\times$31~cm$^{2}$, see Figure~\ref{fig:Sf11_data_muon_xy}). For the DS this is the area between \mbox{$-54\,\textrm{cm}\leq x \leq-2\,\textrm{cm}$} and \mbox{$12\,\textrm{cm}\leq y \leq64\,\textrm{cm}$} (52$\times$52~cm$^{2}$, see Figure~\ref{fig:DS13_data_muon_xy}). 

\subsection{Systematic uncertainties}
The Monte Carlo simulation has shown that there are differences in the tracking performance of ST and HT depending on the energy spectrum (20\% for SciFi and 10\% for DS, see~\cite{Ilieva:2859193}). Because of this dependence, the choice of tracking method introduces a bias in the flux result as it gives preference to muons in certain energy bins. However, the energy spectrum of the data is unknown, and hence the bias can not be determined. For this reason, the difference of the muon fluxes obtained for tracks built with the ST and HT methods is assigned to a systematic uncertainty.

Since the tracking efficiency directly enters the muon flux estimate, its uncertainty is assigned to a systematic uncertainty.

The third source of systematic uncertainty is the integrated luminosity for an LHC~fill, whose value is used to normalize the muon flux. The ATLAS collaboration reports a 2.2$\%$ uncertainty in the integrated luminosity for data recorded in 2022~\cite{ATL-DAPR-PUB-2023-001}. 

The systematic uncertainties per source are given in Table~\ref{tab:systematics} for the SciFi and the DS. For the SciFi the dominant source of uncertainty is the choice of tracking method, while for the DS muon detector it is the tracking efficiency. The total systematic uncertainty is the quadrature sum of the uncertainties for all sources.

\begin{table}[t]
   \centering
    \caption{Relative magnitude of the sources of systematic uncertainty for the  muon flux measurement: luminosity, fluctuations of the tracking efficiency in different $x-y$ detector regions, and the choice of tracking method.}\label{tab:systematics}
    \begin{tabular}{c|c|c|c}
        \hline \hline
         system &\makecell{luminosity\\uncertainty\\$[\%]$}  & \makecell{tracking\\ efficiency\\$[\%]$} & \makecell{choice of\\ tracking method\\$[\%]$}\\
         \hline
         SciFi & 2.2 & 2.2 & 4.8\\
         DS & 2.2 & 2.9 & 2.0\\
         \hline \hline
    \end{tabular}
\end{table}

\subsection{Results}

The  muon flux per integrated luminosity for SciFi and DS are presented in Table~\ref{tab:final_muon_rates_data}, together with the statistical and systematic uncertainties. The DS muon flux is larger than the SciFi  flux because of the non-uniform distribution of tracks in the vertical direction (see Figures~\ref{fig:Sf11_data_muon_xy}~and~\ref{fig:DS13_data_muon_xy}) and the difference in acceptance. The total relative uncertainty is 6~$\%$ for the SciFi measurement and 4~$\%$ for the DS.

\begin{table}[t]
   \centering
    \caption{Muon flux in the SciFi and the DS detectors.}\label{tab:final_muon_rates_data}
    \begin{tabular}{ccc|ccc}
        \hline \hline
         & system &&&muon flux [10$^{4}$ fb/cm$^{2}$]&\\
         \hline
         & SciFi &&& $2.06\pm0.01(\textrm{stat.})\pm0.12(\textrm{sys.}) $ &\\
         & DS &&& $2.35\pm0.01(\textrm{stat.})\pm0.10(\textrm{sys.}) $ &\\
         \hline \hline
    \end{tabular}
\end{table}

\section{Muon flux in the emulsion}\label{sec:emulsion}
During the commissioning phase of the LHC, a reduced target was instrumented with a single brick to establish whether the occupancy of the emulsion could be determined, thus providing input for the analysis of future targets. 
\subsection{Detector layout and track reconstruction}\label{subsec:emulsion_detector}

\begin{figure}[htbp]
    \centering
    \includegraphics[width = 0.4\textwidth]{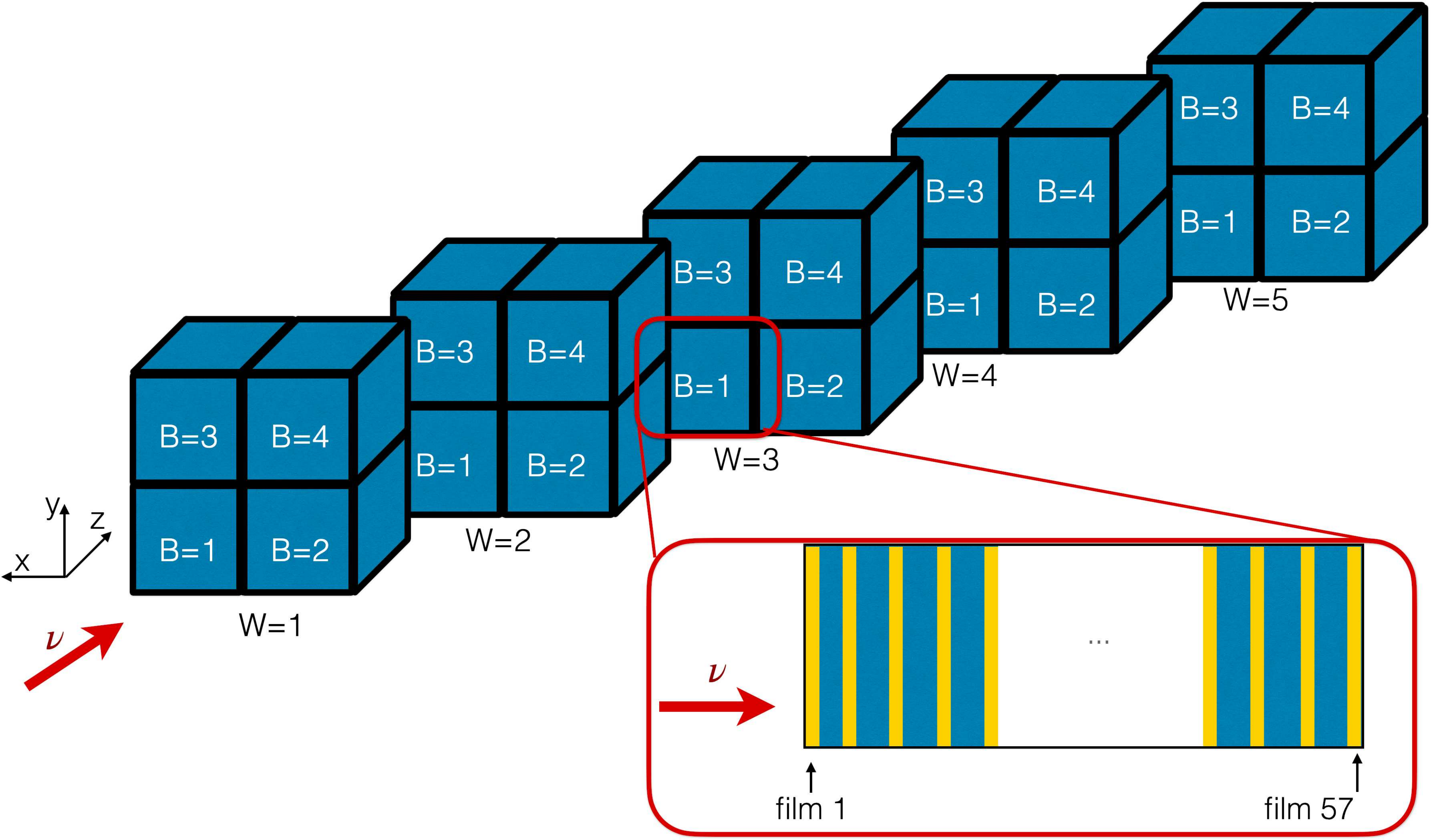}
    \caption{Layout of the \SND emulsion target. The brick instrumented emulsions is highlighted, showing a layout of its inner structure along its depth.}
    \label{fig:emulsion_layout}
\end{figure}

Figure~\ref{fig:emulsion_layout} shows the layout of Emulsion Target 0 that took the data used in this analysis. The ECC brick was located in the third wall, in the position closest to the line of sight in the transverse plane. 

After development and scanning, the data reconstruction was performed with the FEDRA~ROOT~C++ library~\cite{Tyukov:2006ny}. The films were aligned in the reference system of the ECC brick and the tracks in the emulsion target were reconstructed~\cite{Iuliano:2868917}.

\subsection{Angular distribution }\label{subsec:emulsion_results}

\begin{figure}[htbp]
    \centering
    \includegraphics[width = 0.37\textwidth]{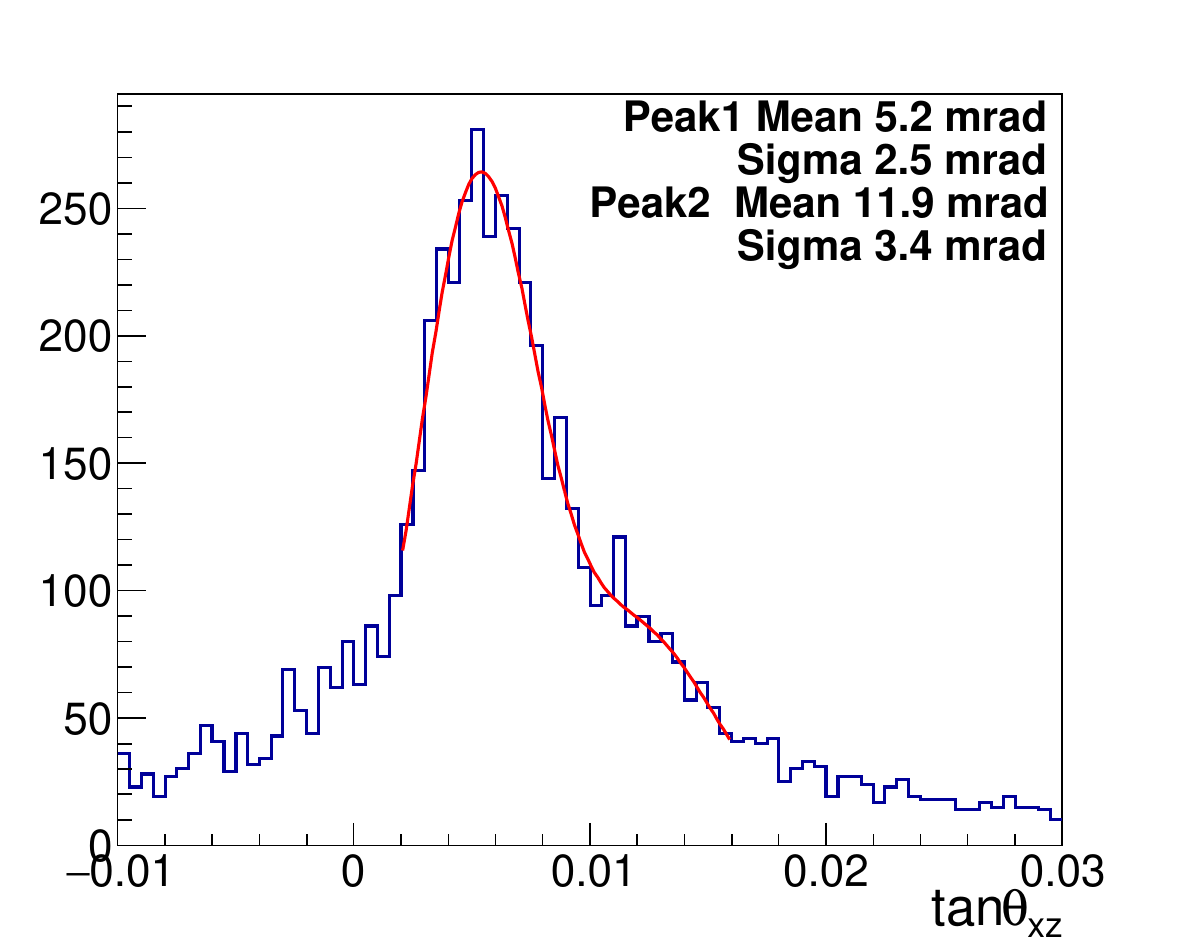}
    \caption{Angles of emulsion tracks in the $x-z$ plane. The two peaks are fitted with a two Gaussian function.}
    \label{fig:track_slopes}
\end{figure}

The angular distribution of the tracks reconstructed in the emulsion is shown in Figure~\ref{fig:track_slopes}. The presence of a second peak in the $x-z$ component can be seen. Fitting the distribution of the angle in the $x-z$ plane with a two Gaussian function results in a distance between the peaks of $\sim$~\qty{6}{mrad}. The angular distribution with two peaks is similar to that observed with the electronic detectors in Figure~\ref{fig:central_peaks}.

\subsection{Result}
\begin{figure}[htbp]
    \centering
    \includegraphics[width = 0.35\textwidth]{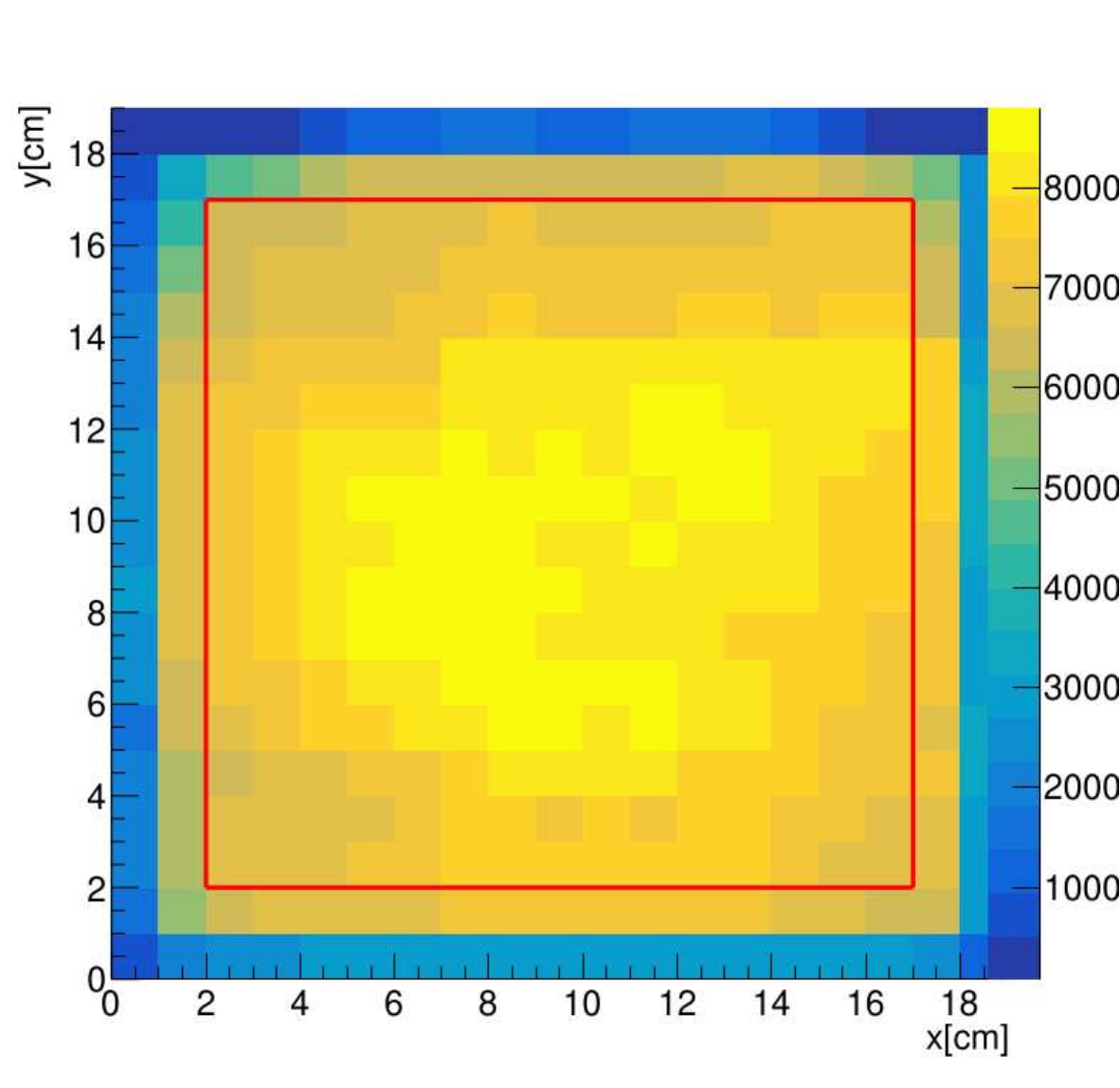}
    \caption{Distribution of tracks at the most upstream film in each \qty{1}{\square\centi\metre} cell, corrected for reconstruction efficiency. The red border represents the region considered for measuring the average density. The coordinates on the axes are local coordinates on the surface of the brick.}
    \label{fig:emu_track_density}
\end{figure}

The spatial density of the reconstructed tracks, after correcting for the tracking efficiency~\cite{Iuliano:2868917}, is shown in Figure~\ref{fig:emu_track_density}. In the region represented within the red border, the  track density is \mbox{$7.7 \pm 0.6~(\textrm{stat})\times 10^3$\,~cm$^{-2}$}. For the luminosity integrated in the emulsion target during the exposure time, the track density corresponds to a  muon flux of \mbox{$1.5 \pm 0.1~(\textrm{stat}) \times 10^4$\,fb/cm$^{2}$}.

\section{Cross checks}\label{sec:crosschecks}
As a cross check of the dependence of the muon flux on the acceptance, the muon flux was also estimated using the SciFi $x-y$ region as acceptance for the electronic subdetectors (see Table~\ref{tab:muFlux_detComp}). In this case, the relative difference between the two measurements is~2$\%$.

The muon flux measured with the SciFi detector is higher than the result obtained from the analysis of the ECC brick ( see Section~\ref{subsec:emulsion_results}). This is due to the vertical gradient of the flux. In order to perform a reliable comparison, the data from the SciFi in the same region in the transverse plane of the ECC was used for analysis. The resulting muon flux is \mbox{$1.6 \pm 0.01 (\textrm{stat}) \pm 0.10 (\textrm{sys})$\,fb/cm$^{2}$}, consistent with the measurement obtained from emulsion.

\begin{table}[htbp]
     \centering
    \caption{Muon flux in the SciFi and the DS detectors in identical detector areas: \mbox{$-42\,\textrm{cm}\leq x \leq-11\,\textrm{cm}$} and \mbox{$18\,\textrm{cm}\leq y \leq49\,\textrm{cm}$}.} \label{tab:muFlux_detComp}
    \begin{tabular}{ccc|ccccc}
        \hline \hline
         &\makecell{system} &&& \makecell{muon flux [10$^{4}$ fb/cm$^{2}$]\\ \textit{same fiducial area}}& \\
         \hline
         &SciFi &&& $2.06\pm0.01(\textrm{stat.})\pm0.12(\textrm{sys.})$& \\
         &DS &&& $2.02\pm0.01(\textrm{stat.})\pm0.08(\textrm{sys.})$& \\
         \hline \hline
    \end{tabular}
\end{table}

\section{Monte Carlo simulation expectation}\label{subsec:MC_comp}

\begin{figure}[htbp]
    \centering
     \includegraphics[width=0.37\textwidth]{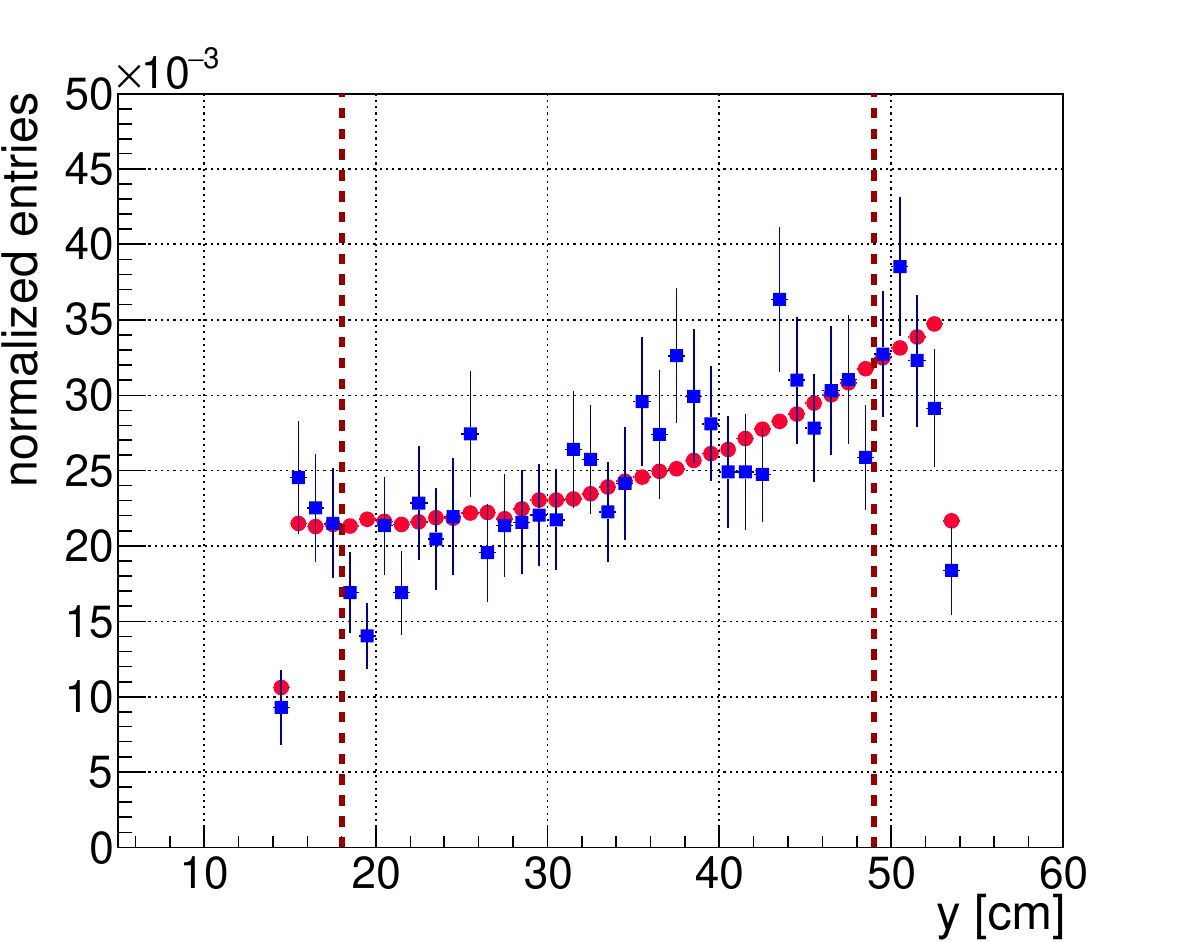}
    \caption{SciFi tracks in data (red) and Monte Carlo simulation (blue) as a function of $y$. The dotted red lines indicate the  $y$ coordinate boundaries of the detector regions selected for the flux estimation. Each distribution is normalized to unit integral.}
    \label{fig:Scifi_data_MC_y}
\end{figure}

\begin{figure}[htbp]
    \centering
     \includegraphics[width=0.37\textwidth]{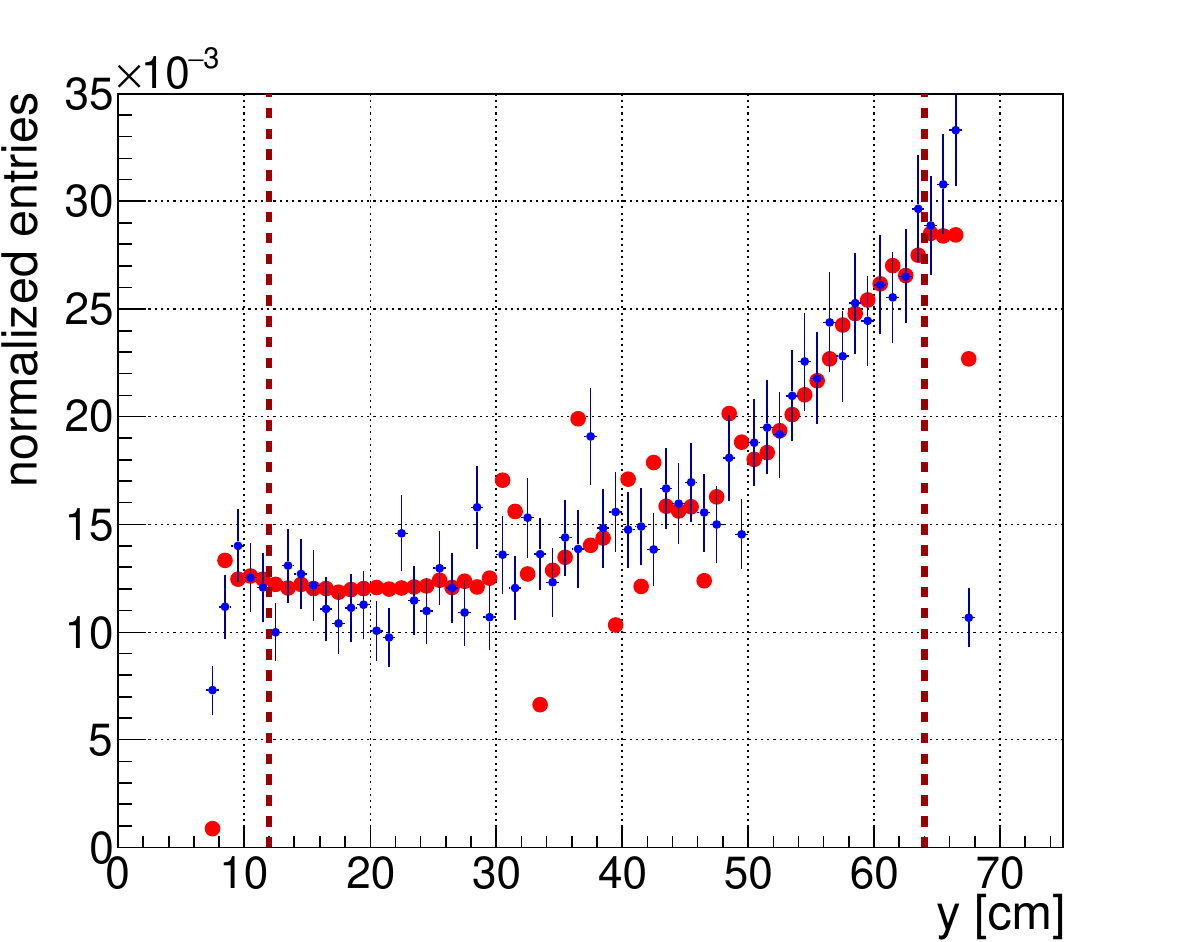}
    \caption{DS tracks in data (red) and Monte Carlo simulation (blue) as a function of $y$. The dotted red lines indicate the  $y$ coordinate boundaries of the detector regions selected for the flux estimation. Each distribution is normalized to unit integral. }
    \label{fig:DS_data_MC_y}
\end{figure}

The non-uniform distribution of tracks in the vertical direction in data (see Figures~\ref{fig:Sf11_data_muon_xy} and \ref{fig:DS13_data_muon_xy}) is also  present in the Monte Carlo simulation (see Figures~\ref{fig:Scifi_data_MC_y} and \ref{fig:DS_data_MC_y}). This is due to the complex magnetic field in the LHC. The larger fluctuations in the simulation are due to limited statistics.  The few outlier data points in the DS (see Figure~\ref{fig:DS_data_MC_y}) are due to inefficient bars (see also Figure~\ref{fig:DS13_data_muon_xy}). 

\begin{figure}[htbp]
    \centering
     \includegraphics[width=0.37\textwidth]{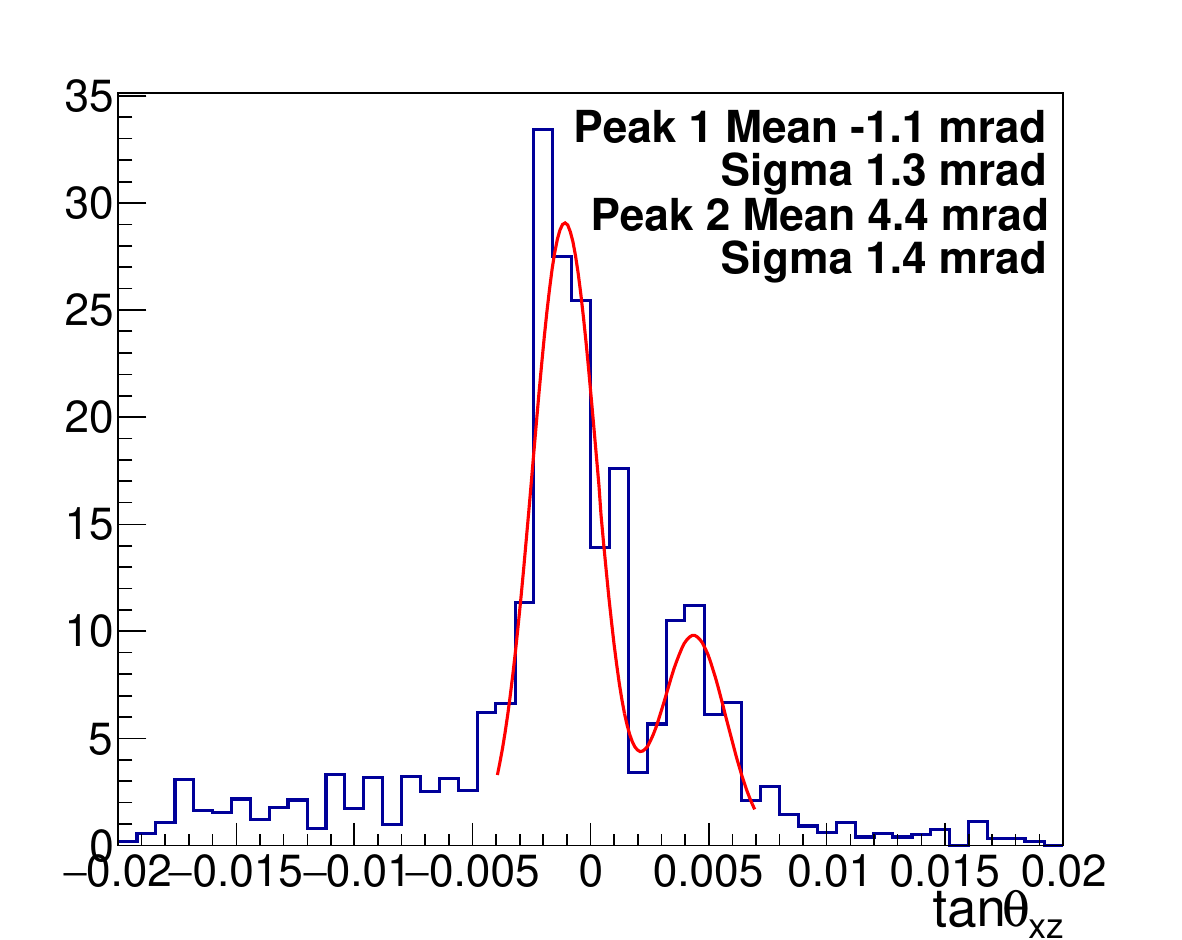}
    \caption{Simulated SciFi track slopes in the horizontal plane. The region around a few~milliradian shows two peaks, which are fitted with a two Gaussian function. }
    \label{fig:MC_central_peaks}
\end{figure}

The two peaks in the angular distribution of the tracks observed in data (see Figures~\ref{fig:central_peaks} and \ref{fig:track_slopes}) are also visible in the MC simulation at a distance of 5.5~mrad (see Figure~\ref{fig:MC_central_peaks}). 

The flux values obtained from the electronic detectors using data  are  between 20--25~$\%$ larger than those obtained from the Monte Carlo simulation (see  Table~\ref{tab:MC_data_comparison}).


There are many physics processes underlying the production of muons from $pp$ collisions,  the production of muons through
decays of non-interacting pions and kaons, as well as their transportation  through magnetic elements of the LHC and several hundred meters of rock. Given this complexity and the fact that the three stages are simulated with different Monte Carlo programs, each with an associated uncertainty ranging from 10$\%$--200$\%$, the agreement between the prediction by the Monte Carlo simulation and the measured flux is remarkable.


\begin{table}[t]
    \caption{Comparison between the muon flux obtained from data and Monte Carlo simulation.} \label{tab:MC_data_comparison}
    \centering
    \begin{tabular}{c|c|c|c}
        \hline \hline
         &&\\
         \hspace{-0.12cm}system\hspace{-0.12cm} & \hspace{-0.13cm}sample\hspace{-0.13cm} &
         \makecell{muon flux\\ $[10^{4}$ fb/cm$^{2}]$} & \hspace{-0.27cm}\makecell{1$-$$\frac{\textrm{sim}}{\textrm{data}}$\vspace{0.07cm} \\$[\%$]}\hspace{-0.27cm}\\
         &&\\
         \hline
         SciFi & \makecell{data\\sim} & \makecell{$2.06\pm0.01(\textrm{stat.})\pm0.12(\textrm{sys.}) $\\ $1.60\pm0.05(\textrm{stat.})\pm0.19(\textrm{sys.})$} & 22 $\pm$ 9\\
         DS & \makecell{data\\sim} & \makecell{$2.35\pm0.01(\textrm{stat.})\pm0.10(\textrm{sys.})$\\ $1.79\pm0.03(\textrm{stat.})\pm0.15(\textrm{sys.})$ } & 24 $\pm$ 9\\
         \hline \hline
    \end{tabular}
\end{table}

\section{Conclusion}\label{sec:conclusions}

The muon flux at \SND  is measured using  three independent tracking detectors: ECC, SciFi and the DS. The analyzed data samples were taken during the 2022 LHC proton run. The muon flux per integrated luminosity through an \mbox{18$\times$18 cm$^{2}$} area of one ECC brick is \mbox{$1.5 \pm 0.1(\textrm{stat}) \times 10^4\,\textrm{fb/cm}^{2}$}. The measured muon flux using the SciFi and a \mbox{31$\times$31 cm$^{2}$} area between \mbox{$-42\,\textrm{cm}\leq x \leq-11\,\textrm{cm}$} and \mbox{$18\,\textrm{cm}\leq y \leq49\,\textrm{cm}$} is \mbox{$2.06\pm0.01(\textrm{stat})\pm0.12(\textrm{sys}) \times 10^{4}\,\textrm{fb/cm}^{2}$}. With the DS muon detector and a \mbox{52$\times$52 cm$^{2}$} area, the muon flux is \mbox{$2.35\pm0.01(\textrm{stat})\pm0.10(\textrm{sys}) \times 10^{4}\,\textrm{fb/cm}^{2}$.} The difference between the estimates is due to a vertical gradient of the flux and the different areas of acceptance of SciFi and DS. The total relative uncertainty of the electronic detectors results is 6~$\%$ for the SciFi and 4~$\%$ for the DS measurement. When considering the same area of acceptance for  SciFi and  ECC, or for  SciFi and  DS, the measured muon fluxes are in good agreement.

\section{Acknowledgements}
We express our gratitude to our colleagues in the CERN accelerator departments for the excellent performance of the LHC. We thank the technical and administrative staffs at CERN and at other SND@LHC institutes  for their contributions to the success of the SND@LHC effort. 
We acknowledge and express gratitude to our colleagues in the CERN SY-STI team for the fruitful discussions regarding beam losses during LHC operations. 
In addition, we acknowledge the support for the construction and operation of the SND@LHC detector provided by the following funding agencies:  CERN;  the Bulgarian Ministry of Education and Science within the National
Roadmap for Research Infrastructures 2020–2027 (object CERN); ANID—Millennium Program—$\rm{ICN}2019\_044$ (Chile); the Deutsche Forschungsgemeinschaft (DFG, ID 496466340); the Italian National Institute for Nuclear Physics (INFN); JSPS, MEXT, the Global COE program of Nagoya University, the Promotion
and Mutual Aid Corporation for Private Schools of Japan for Japan;
the National Research Foundation of Korea with grant numbers 2021R1A2C2011003, 2020R1A2C1099546, 2021R1F1A1061717, and 
2022R1A2C100505; Fundação para a Ciência e a Tecnologia, FCT (Portugal), 
CERN/FIS-INS/0028/2021; the Swiss National Science Foundation (SNSF); TENMAK for Turkey (Grant No. 2022TENMAK(CERN) A5.H3.F2-1).
M.~Climesu, H. Lacker and R.~Wanke are funded by the Deutsche Forschungsgemeinschaft (DFG, German Research Foundation), Project 496466340. We acknowledge the funding of individuals by Fundação para a Ciência e a Tecnologia, FCT (Portugal) with grant numbers  CEECIND/01334/2018, 
CEECINST/00032/2021 and PRT/BD/153351/2021. We thank Luis Lopes, Jakob Paul Schmidt and Maik Daniels for their help during the construction. 

\bibliography{sn-bibliography}

\end{document}